\documentclass[]{aa}
\usepackage{graphics}
\usepackage{txfonts}
\usepackage{natbib}
\begin{document}

\title
{Tracing a Disk Wind in NGC 3516 
}
\subtitle{}

\titlerunning{A Disk Wind in NGC 3516}

\author
{T.\ J.\ Turner\inst{1,2} \and 
J.\ N.\ Reeves\inst{3} \and 
S. \ B.\ Kraemer\inst{4} \and 
L.\ Miller\inst{5} 
}

\authorrunning{T.\ J.\ Turner et al.\ }

\institute{Dept. of Physics, University 
of Maryland Baltimore County, 1000 Hilltop Circle, Baltimore, MD 21250
\and 
Code 662,  Exploration of the Universe Division,   
NASA/GSFC, Greenbelt, MD 20771
\and 
Astrophysics Group, School of Physical and Geographical Sciences, 
Keele University, Keele, Staffordshire ST5 5BG, UK
\and
Department of Physics, Catholic University of America, 620 Michigan 
Avenue NE, Washington, DC 20064
\and
Dept. of Physics, University of Oxford, 
Denys Wilkinson Building, Keble Road, Oxford OX1 3RH, U.K.
}

\date{Received / Accepted}

\abstract
{X-ray spectra of AGN often contain signatures indicative of absorption in multiple 
layers of gas whose ionization-state and covering fraction may vary with time. It has been unclear to date how 
much of the observed X-ray  spectral and timing behavior in AGN can be attributed to variations in 
absorption, versus 
variations in the strengths of emission or reflection components. Diagnostics of the inner 
regions of AGN cannot be reliably performed until the origin of observed effects is understood. }
{We investigate the role of the  X-ray absorbers in the Seyfert 1 galaxy NGC 3516}  
{Time-averaged and flux-selected spectroscopy is used to examine the behavior of NGC 3516 observed in 
Chandra HETG and XMM data from Oct 2006. }
{New H-like and He-like emission and absorption features discovered in the Fe K regime 
 reveal a previously unknown zone of circumnuclear gas in NGC 3516 with log $\xi \sim 4.3$ 
and column density  $\sim 10^{23} {\rm cm^{-2}}$. A lower-ionization layer with log $\xi \sim 2$ and of similar column 
density  is confirmed from previous observations, this layer has a  covering fraction around 50\%, and changes in covering 
provide a simple explanation of a deep dip in the light curve that we interpret as  an eclipse of the continuum 
due to passage of a cloud across the sight line within half a day. 
These inner zones of absorbing gas are detected to have outflow velocities in the 
range $1000-2000$\,km\,s$^{-1}$, this, and constraints 
on radial location are consistent with an origin as part of a  disk wind in NGC 3516. }
{}

\keywords{accretion; galaxies: active}

\maketitle

\section{Introduction}

The process of accretion onto a supermassive black hole feeds the so-called 'active' subset of galaxies.  
The centers of galaxies are regions where material must exist in the strong gravity regime and 
in active galaxies there is also abundant radiation emerging in the X-ray bandpass from that region.  
The interaction of nuclear X-rays with circumnuclear gas 
allows us a direct probe of the extreme conditions close to an event horizon. 
It is widely thought that the large amount of angular momentum that accreting material must have means 
this material forms an `accretion disk'  around the black hole,  extending to within a few gravitational radii  
(the innermost stable orbit depending on the angular momentum of the black hole). 
Other significant components likely exist  in the nuclear environs including a corona of high energy particles 
sandwiching the disk, a wind of material ablated from the disk surface and clouds of gas at various radial locations. 
\citet{ckg} review the details of X-ray reprocessors along with estimates of the locations and conditions in various 
distinct zones. 

Intrinsically narrow emission  lines 
arising from the surface of the accretion disk or in any of the gas zones very close to the black hole 
 will be distorted by 
a combination of Doppler and relativistic effects whose details can yield conditions and kinematics of the gas 
(see \citealt{fab} for a review of effects). 
X-ray spectroscopy provides a good probe of gas very close to the black hole as 
Fe K$\alpha$ is a strong emission line, emitted in the X-ray band via 
fluorescence or recombination processes  between  6.4  -- 7 keV (depending on the  
ionization-state of the gas); this is a strong line owing to the high abundance and 
 fluorescence yield of Fe. The high ionization potentials of the highest ionization 
states of iron mean that this line can be  produced 
in regions where the gas is extremely ionized. 

Fe K$\alpha$ emission is commonly observed in AGN and much has been made of the 
diagnostic potential of the line, with many authors deriving accretion disk parameters 
from the line profile with important implications \citep{tanaka,n97}. 
However, spectroscopy at the modest energy resolution currently available in the X-ray band has 
been plagued with ambiguity; complex absorption from ionized gas can produce curvature 
in the 2-6 keV band that looks similar to that of an Fe emission line produced close to the 
event horizon of the accretion disk.  
Recent spectra from {\it XMM} have shown absorption lines from highly ionized ions in several 
AGN, allowing construction of detailed absorption models in some cases and revealing hitherto 
unknown zones of absorbing gas (e.g. \citealt{reeves04}). 
In the broad-line Seyfert 1 galaxy (BLSy1) NGC 3516 (z=0.008836; \citealt{k96}), absorption lines from 
a broad range of ions showed  three distinct zones of gas covering the active 
nucleus in that case and producing much of the spectral curvature and variability in the 
X-ray band \citep{turner05}.
The classic $U$ version of ionization parameter was utilized by \citet{turner05} 
Here we use the $\xi$ form of ionization parameter where  
$\xi = L/nr^2$.       $L$ is the 1-1000 Rydberg luminosity, 
$n$ the gas density and $r$ the absorber-source distance. 
We quote $\xi$ in units of erg\,cm\,s$^{-1}$ throughout and {\sc xstar} tables are based on 
runs with micro-turbulence $\sigma=200$ km\,s$^{-1}$  unless otherwise noted. 
Ionization states found for the the gas zones present during 2001 observations can be converted 
into  $\xi$ values assuming an illuminating intrinsic X-ray continuum 
$\Gamma \sim 2$ with $\alpha_{o-x} \sim 1.5$.  Thus the  2001  results correspond to: 

\begin{itemize}
\item  a cool `UV absorber' with 
log $\xi_1 \sim -0.5$,  $N_H \sim 6 \times 10^{21} {\rm cm^{-2}}$ 

\item an outflowing zone with log $\xi_2 \sim 3$, $N_H \sim 10^{22} {\rm cm^{-2}}$ and outflow 
 velocity $\sim 1100$ km\,s$^{-1}$   

\item a `heavy' absorber with  log $\xi_3 \sim 2.5$, $N_H \sim 10^{23} {\rm cm^{-2}}$ 
covering $\sim 50\%$ of the continuum. 

\end{itemize}

\citet{turner05} found the spectral variability observed within 2001 to be consistent with the absorbing gas layers 
responding in ionization-state as the  continuum flux varied.

\citet{markowitz07} presented a 2005 {\it Suzaku} observation where two zones of 
gas were detected: the primary absorber is a layer of low-ionization 
log $\xi \sim 0.3$ and $N_H \sim 6 \times 10^{22} {\rm cm^{-2}}$ covering $\sim 96-100$\% of the source, while evidence is 
also found for a zone having 
log $\xi \sim 3$ and $N_H \sim 4 \times 10^{22} {\rm cm^{-2}}$. It is unclear 
which of these zones correspond to particular  layers isolated in  2001 data. {\it Suzaku} data were consistent with the presence 
of the UV-absorber detected during 2001, but if we want to associate the primary absorber detected by {\it Suzaku} with the 
outflowing layer detected in 2001, then the gas has undergone a major change in ionization-state between the two epochs; 
if we compare the {\it Suzaku} primary absorber with the 'heavy' layer detected during 2001, then that layer would 
have had to have changed 
$\xi$, $N_H$ and covering fraction.   Whatever the correspondence between modeled zones 
 {\it Suzaku} data clearly show NGC 3516 to have much larger absorption by relatively low-ionization gas 
compared to any of the 2001 data \citep{markowitz07}.  
During 2006 {\it Suzaku} found NGC 3516 to have hard-band flux 
(above $\sim 6 $ keV) 
comparable to that observed during April 2001 and thus the change in absorption seems likely to be 
due to the passage of a relatively cool cloud into the sight-line,  as seen  in X-ray observations of some other 
AGN (e.g. NGC 4151 \citealt{k05}, NGC 1365 \citealt{risaliti07}). 
At this point one could  only conclude that the inner zones of the absorber in 
NGC 3516 change dramatically 
over time in all senses.

X-ray observations of  NGC 3516 also provided the first evidence for 
'transient' narrow emission lines, most easily explained as Doppler shifted Fe emission \citep{turner02,iw}. 
With all of these new discoveries our picture of the 
contributions from various reprocessors is rapidly evolving.

In this paper we present data from a 
coordinated observation of NGC 3516 utilizing {\it Chandra} and {\it XMM} data, exploring further  the 
innermost  absorbing gas in this source.

\section{The Observations}

\begin{figure}[h]
  \resizebox{8cm}{!}{ 
  \rotatebox{0}{
  \includegraphics{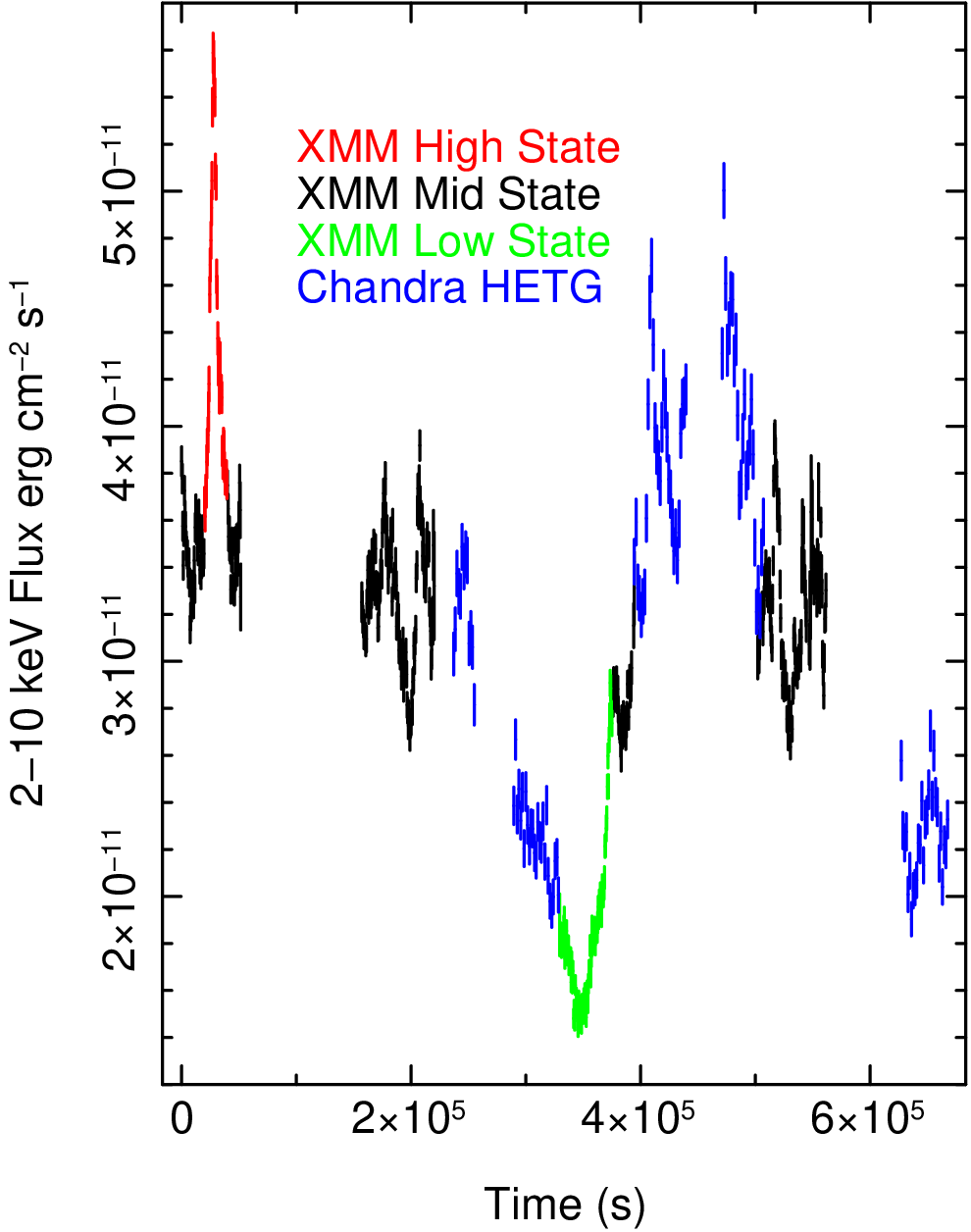}}}
  \caption{XMM pn data in 500 s bins, and summed HEG and MEG 
 data from all orders over 1-6 keV in 1500 s bins. The light curve 
shows the coverage of NGC 3516 with each satellite. 
The XMM data were time-split for spectral analysis, and the red points denote the time selection 
defining the high-flux state; the black points show the medium or mid-flux state, the green points 
show the dip event, comprising the low-flux state} 
\label{lc}
\end{figure}

An {\it XMM-Newton} (hereafter {\it XMM}) observation of NGC 3516 was 
made covering 2006 Oct 6-13 (OBSIDs 04012010401, 501, 601, 1001). {\it Chandra} observations 
(OBSIDs 7281,7282, 8450, 8451, 8452) were interwoven with 
the {\it XMM} exposures, to provide complete coverage of the source during that period. 
Fig.\,\ref{lc} shows the light curve from the two satellites, illustrating the coverage 
obtained and the source behavior across the observation. 

EPIC data utilized the thin1 filter and pn  PrimeSmallWindow mode.   
Data were processed using SAS {\sc v7.0} 
using standard criteria and removing periods of high background. 
 Instrument patterns 
0--4 were selected for the pn and here our analysis describes 
only the pn data, as these  
offer the best S/N ratio of all EPIC data, and were completely free of 
photon pileup effects.  Data were extracted 
from a circular 
cell of radius $40''$ centered on the source, and 
background data were taken from a source-free region of the same pn chip.
Periods where the full-band pn count rate exceeded 2 count\,s$^{-1}$  in 
the background region were excluded from the analysis. 
The combined screening yielded a total effective EPIC pn exposure of 155\,ks 
over an observational baseline of about one week.

\begin{figure}[h]
  \resizebox{8cm}{!}{ 
  \rotatebox{0}{
  \includegraphics{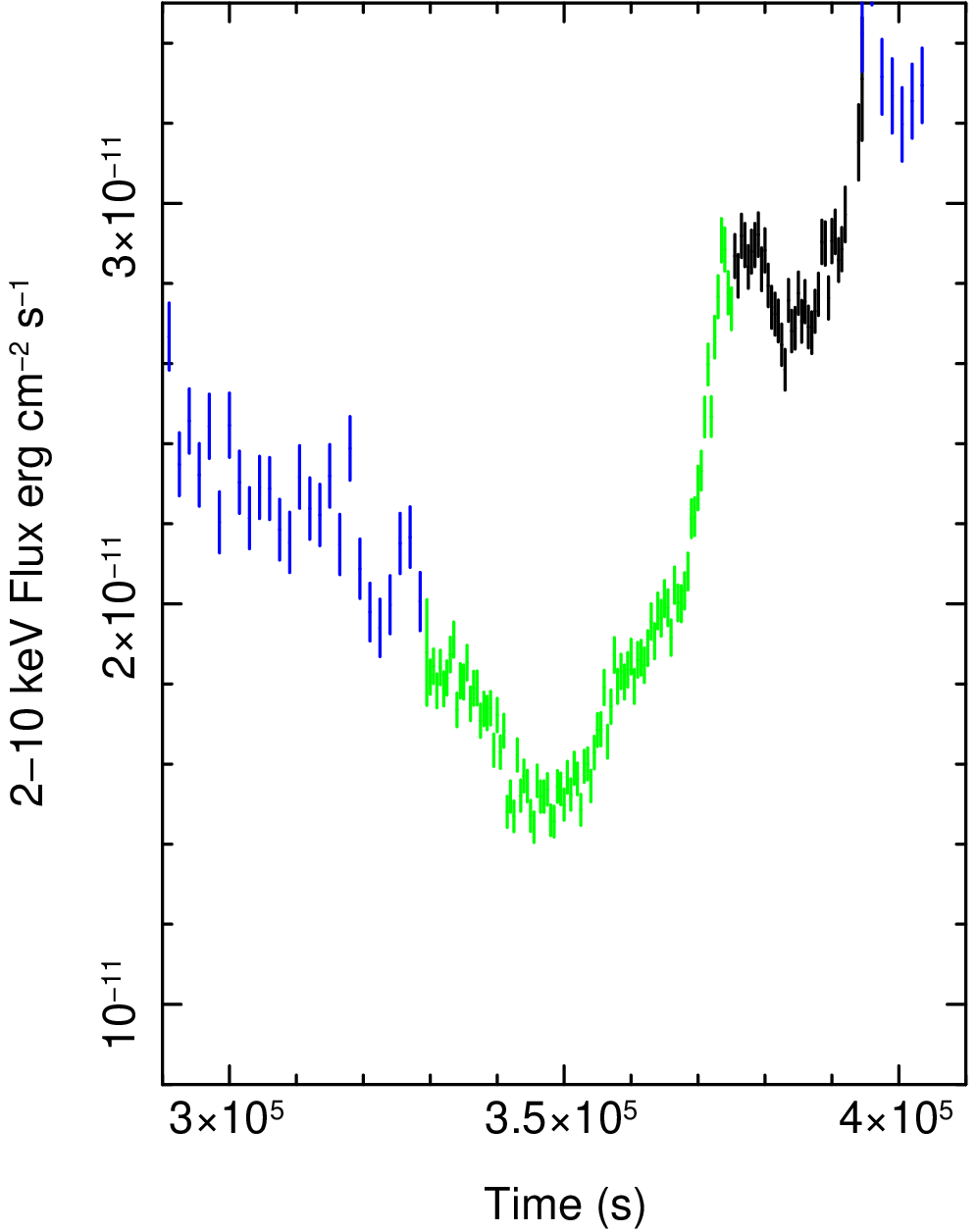}}}
  \caption{A close-up view of the dip event from Fig.~1.} 
\label{dip_zoom}
\end{figure}

NGC 3516 gave a mean pn count rate $\sim 4.91$ $\pm0.007$ count\,s$^{-1}$  in the 2-10 keV band. 
The mean background level in the screened data was $\sim 1$\%  of the 
mean source rate in this band. As evident in Fig.\,\ref{lc}, the source ranges 
across a 2-10 keV flux 
$F \sim 1.4 - 5.4  \times 10^{-11}{\rm erg\ cm^{-2}s^{-1}}$ during the 
course of the observations.  First-order 
RGS data yielded 0.450$\pm 0.001$ and 0.389$\pm 0.001$  count\,s$^{-1}$ in RGS 1 and 2, respectively. 
The RGS background level was $\sim 12\%$ of the total count rate.     

\begin{figure}[h]
  \resizebox{8cm}{!}{ 
  \rotatebox{0}{
  \includegraphics{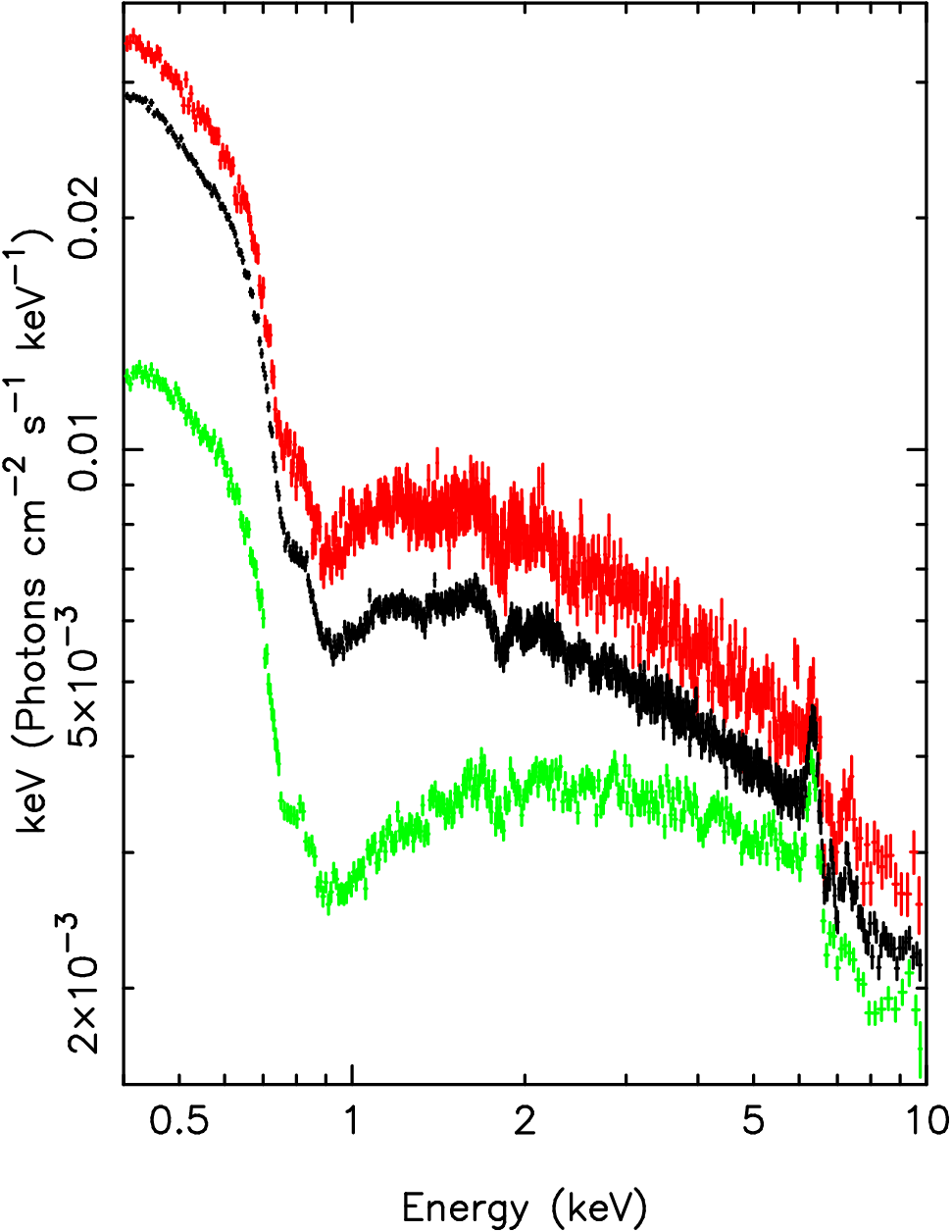}}}
  \caption{X-ray spectra of NGC 3516 from {\it XMM} 2006 corresponding to 
the flux states denoted in Fi.~1, i.e. spectra  are from the 
high/flare state (top), the mid-flux state (middle) and dip state (bottom). 
}
\label{2006_states}
\end{figure}

The {\it Chandra} HETG exposure yielded a 2-7 keV count rate 
$\sim 0.021 \pm 0.0005$ count\,s$^{-1}$  in the summed HEG first order spectrum. 
Data were reduced using CIAOv3.4 and CALDB v3.4.0
 and following standard procedures for extraction of HETG spectra, with the exception that 
we utilized a narrower extraction strip than the {\sc tgextract} default. The default 
processing criterion  
cuts off the HEG data above 8 keV, yet some of the brighter and/or well-exposed sources still have good 
HEG data above 8 keV, including this well exposed observation of NGC 3516.  
This reason for the default processing cut-off is that the overlap of the MEG and HEG strips 
depends 
on the extraction strip widths, and if the latter are too large, a larger intersection of the 
MEG and HEG strips results, cutting off the HEG data prematurely. 
Specifically, we used
\verb+width_factor_hetg=20+ in the tool 
\verb+tg_create_mask+, instead of the default value of 35. As the HETG spectra  have a very low 
signal-to-noise ratio it was necessary to coadd the positive and 
negative first-order spectra, and coadd all five OBSIDs, to create a high quality summed first-order HEG 
and MEG spectra for fitting. 
After such co-addition, we binned the spectra to 4096 channels, which yielded $>20$ counts per bin 
in the fitted range.

\begin{figure}[h]
  \resizebox{8cm}{!}{ 
  \rotatebox{0}{
  \includegraphics{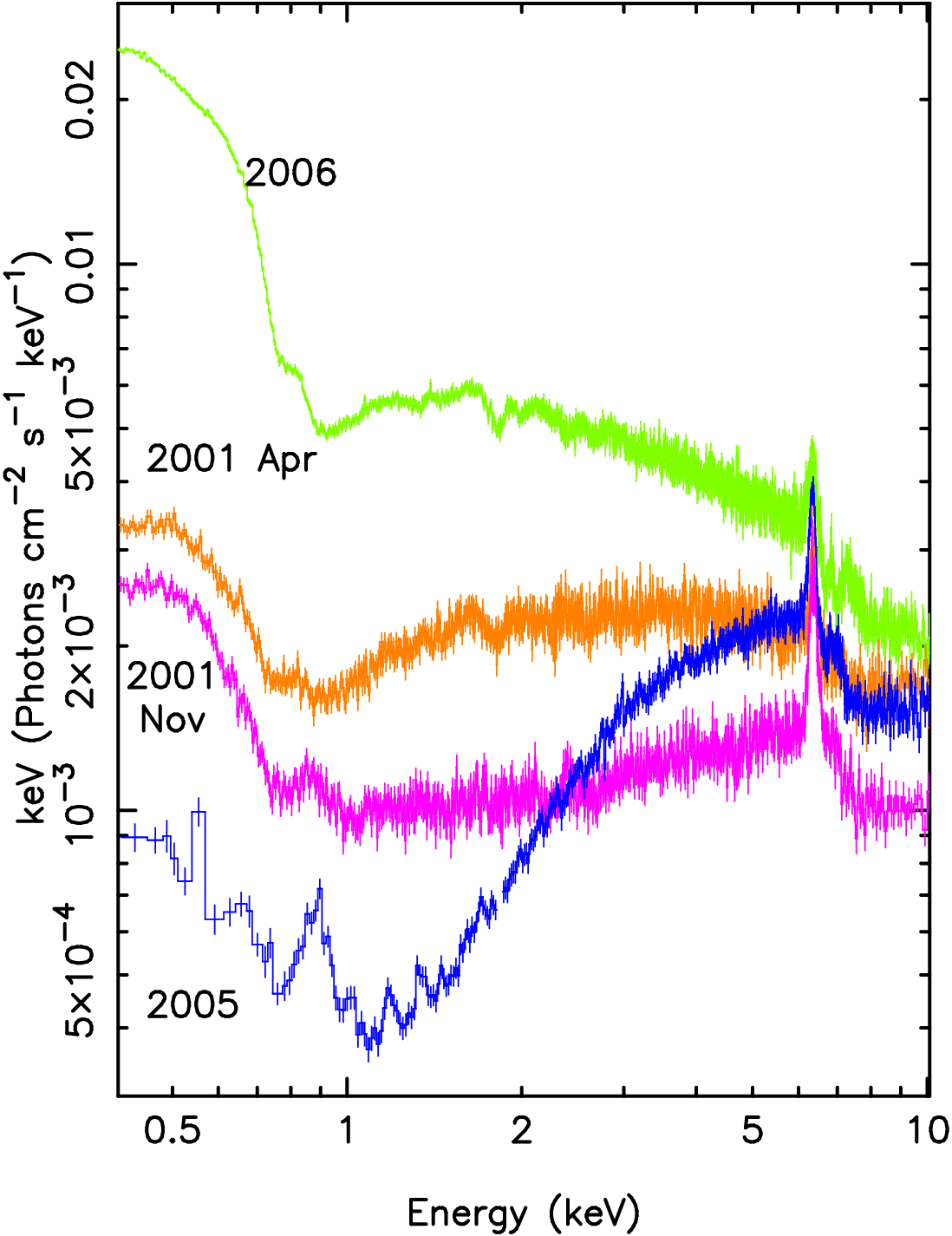}}}
  \caption{X-ray spectra of NGC 3516 from {\it XMM} 2001 April,  2001 Nov, 
 {\it Suzaku} data  from 2005 October and 2006 Oct {\it XMM} data} 
\label{all}
\end{figure}

\section{Results}

The light curve from the combined observation shows the 
source to exhibit significant flux 
variability (Fig.\,\ref{lc}), 
with a notable flare early in the observation, and a  drop into a 'dip' state 
later in the observation (Fig.\,\ref{dip_zoom}). Comparison of spectra 
from the different states show a flattening as the source flux drops (Fig.\, \ref{2006_states}) 
as often observed in sources of this class. 
Comparison of the new spectra with  2001 {\it XMM} pn  and  2005 {\it Suzaku} XIS data illustrates the 
relatively high flux level at which the source was observed during 
2006 (Fig.\,\ref{all}).

Inspection of the detailed 
profile of the light curve during the dip reveals interesting 
structure (Fig.\,\ref{dip_zoom}), with the 
source flux levelling off for a period 
lasting $\sim 10$ ks. 
Intriguingly, the initial shallow dip egress has a duration of 20 ks, but breaks 
to a sharp rise 
occuring over just 6 ks.

\begin{figure}
  \resizebox{6cm}{!}
{
  \rotatebox{0}{
  \includegraphics{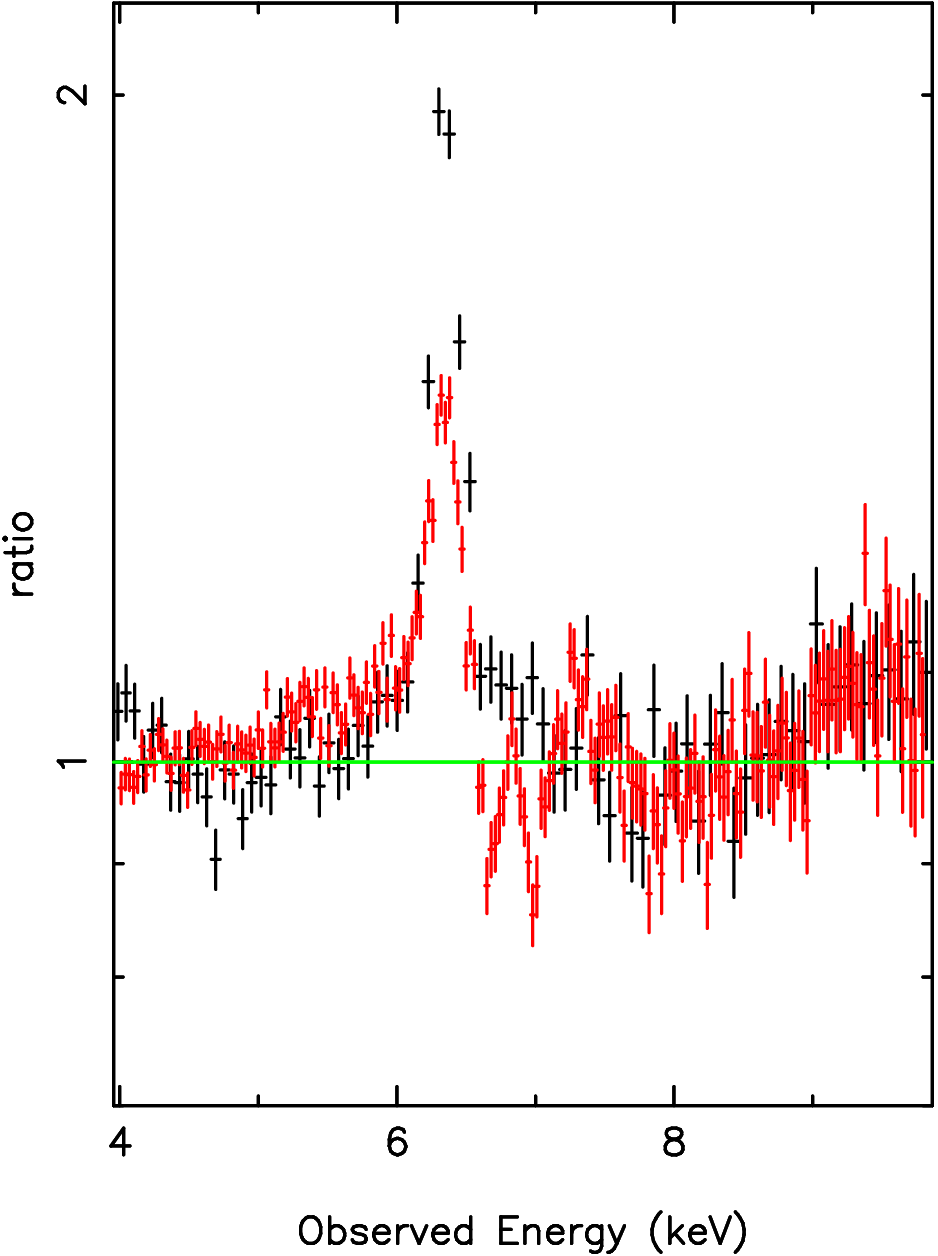}}}
  \caption{{\it XMM} pn data shown as a ratio to the local 
continuum fit in the Fe K regime 
during Oct 2006 (red) and Nov 2001 (black). 
}
\label{absnvar}
\end{figure}

We first analyzed the time-averaged pn spectrum.
 Fig.\,\ref{absnvar} 
shows a fit to the pn data in the Fe K-band with 2006 data shown in red. 
The most obvious result is the presence of two deep absorption lines in the 6.5-7 kev regime, 
naturally identified with He-like and H-like species of Fe. Fitting the line pair with simple gaussian 
profiles we find 
an equivalent width 100$^{+19}_{-12}$ eV for the line at $E=6.989 $ keV, and $45^{+10}_{-7}$ eV for the line at 
$E=6.697$ keV, both measured against the total 'local continuum'. To 
obtain the most accurate line energies and 
widths we turned to the contemporaneous HETG data. 

Fig. \,\ref{heg} shows the  HEG spectrum of NGC 3516 from 2006. The 
deep absorption lines from Fe are confirmed, as are narrow emission lines in the same band. 
Of great interest is the detection of an emission line redward of the H-like Fe absorption feature.  

Fitting the 
emission and 
absorption line `pair' close to 7 keV  yields energies close 
to those of rest-frame H-like Fe transitions; 
the red-shifted emission and 
blue-shifted absorption of these components constitutes a 
profile somewhat akin to a classic `P Cygni' profile,
 indicative of an origin in an outflowing layer of gas, although 
in this case the emission line is 
offset from the expected rest energy, as well as the absorption line. 
 A redshifted emission component 
might be expected if the outflow 
arises in gas that is not a simple expanding shell. A conical outflow, for example, could produce a 
redshifted emission component and a blue-shifted absorption component, if viewed at orientations along the flow axis of the cone. 
 The Fe line components from H-like Fe 
are consistent with an origin in an  outflow of velocity of $\sim 1000$\,km\,s$^{-1}$. 
 The emission component of this H-like Fe  pair 
is too narrow and weak to show up in the {\it  XMM} pn spectra, but the pn spectra taken at approximately the same time and flux-state 
(the mid-state, Fig.\,\ref{lc}) are consistent with the presence of such a component. 

\begin{table*}
{\small
\begin{tabular}{r@{}lr@{}lrrlr@{}ll}
 & Energy  & & $\sigma$ & \multicolumn{1}{c}{$n$} & EW & ID & v  &  & $\Delta \chi^2$ \\ 
 & & & & (10$^{-5}$ photon & &  &  & \\
 & (keV) & & (eV) & cm$^{-2}$\,s$^{-1}$) & (eV) &  & (km\,&s$^{-1}$) &  \\
\hline

1&.3567$\pm 0.002$ & 6&.14$^{+0.79}_{-0.66}$ & $-3.17^{+0.76}_{-0.77}$ & $-8$ & Mg {\sc xi} & $-991$&$\pm440$ & 57 \\

1&.4658$\pm0.007$  & 6&.14 & 0.75$^{+0.52}_{-0.50}$    & 2 & Mg {\sc xii}   & 1302&$\pm1440$ &  13 \\ 

1&.4806$\pm0.001$ & 6&.14  & $-2.87^{+0.73}_{-0.55}$   &   $-8$ & Mg {\sc xii} & $-1710$&$\pm201$ & 160 \\

1&.8709$\pm 0.003$  & 8&.51$^{+1.26}_{-1.18}$  &  $-2.52^{+0.51}_{-0.56}$   &  $-9$  & Si  {\sc xiii} & $-911$&$\pm625$ & 65 \\

1&.999$\pm 0.007$ & 8&.51 & 0.89$^{+0.63}_{-0.57}$   & 3 &  Si {\sc xiv} &  968&$\pm1050$ & 17 \\

2&.018$\pm 0.002$  & 8&.51 & $-3.24^{+0.53}_{-0.66}$   &   $-12$  & Si {\sc xiv}  & $-2001$&$\pm 431$ &  301 \\

2&.640$\pm0.007$ & 11&.4$^{+6.6}_{-5.5}$  & $-1.46^{+0.79}_{-1.14}$  & $-7$ & S {\sc xiv} & $-2591$&$\pm 719$ & 10 \\ 

6&.404$\pm0.019$ & 40&.0$^{+10}_{-15}$ & 5.55$^{+1.97}_{-1.76}$   & 94 & Fe {\sc i} & $-200$&$\pm 690$ & 31 \\

6&.700$\pm0.06$ & 62&.0$^{+35}_{-93} $ & $-3.74^{+1.89}_{-1.81}$  & $-60$ & Fe {\sc xx} &  &  $-$   & 14 \\

6&.940$\pm0.02$ & 16&.0$^{+8}_{-16}$  &  $3.61^{+5.73}_{-1.88}$  & 89 & Fe {\sc xxvi}  & $1297$&$\pm 827$ & 6 \\

6&.990$\pm0.04$ & 54&.0$^{+15}_{-10}$  & $-6.13^{+1.50}_{-4.24}$  & $-112$ &  Fe {\sc xxvi}  & $-860$&$\pm 1654$  
& 22 \\

7&.050 (f) &  \ 40.0 & & \multicolumn{1}{c}{f} & 14 & Fe I K$\beta$ &  &  & 1 \\
\end{tabular}
}
\caption{Spectral Lines detected in HETG data. Fixed parameters are denoted (f)}
\label{scatttble}
\end{table*}

Having tight constraints on line energies and widths from HEG we tested the 2001 {\it XMM} pn and HETG 
data for consistency with the presence of H-like and He-like absorption lines at the same observed 
energies and widths as those measured during 2006. The pn spectra were inconsistent at the 1\% significance level    
with the presence of absorption lines at the same energies, widths and depths as observed in 2006. 
Specifically, the {\it XMM} data limit the presence of H-like Fe absorption to be an order of magnitude 
weaker during Nov 2001 and more than a factor 2 weaker during April 2001 than was observed during 2006. 
Fig.\,\ref{absnvar} shows 2001 November pn spectral data overlaid with that 
from 2006, illustrating the difference in spectral signatures for the two 
epochs. We also analyzed the  {\it Suzaku}  data for NGC 3516. The data 
were reduced following
 the analysis procedure of \citealt{markowitz07} and the XIS 0, 2 and 3 spectra were summed  for spectral fitting. The data were fit 
for the presence of the H-like Fe absorption line and fitting 
 ruled out the presence of a line of the same flux or
equivalent width as observed during 2006, at the 
 1\% significance level (assuming the line energy 
and width to be as detected by HETG). 

We then searched the low energy part of the 2006 HETG data, fitting MEG and HEG data together 
to search for narrow features. Several lines are significantly detected 
(Fig.\,\ref{meg}) 
and were modelled with a simple gaussian form.  
Table~1 lists lines detected across $\sim 1$ - 8 keV in HETG data.  
Line energies tabulated have been  corrected for the systemic redshift 
(z=0.008836).  Photon fluxes are given with positive values 
indicating an emission line and negative an absorption line; 
 equivalent widths are measured relative to the local observed continuum. 
The  velocity implied for the gas is noted for each detected line, where 
negative velocity indicates outflow (no velocity  is given for the 6.7 keV 
line as this is 
 an uncertain blend of ionization states). 
The last column of Table~1 shows the 
 $\chi^2$ value when the line is added to the model. 
Quoted errors are 90\% confidence intervals.  The Fe K$\beta$ line in Table~1 
was scaled to the K$\alpha$ component but was not 
significantly detected in its own right. The K$\beta$ energy was fixed 
appropriate to an origin in neutral gas and constrained  to 
have flux 11.3\% that of the neutral Fe K$\alpha$ line component. 
For groups of lines with closely-spaced energies we linked the line 
widths together in the fit 
so a turbulent velocity could be estimated  based upon several lines 
together; for 
lines whose widths were linked the same value is shown for the 
$\sigma$ entry for each but only the first occurrence in Table~1  
carries the combined error. 

The fitted widths correspond to 
velocity dispersion $\sigma_{Mg} =1245^{+160}_{-138}$\,km\,s$^{-1}$, $\sigma_{Si}=1265^{+187}_{-176}$km\,s$^{-1}$  and 
$\sigma_S=1292^{+751}_{-625}$km\,s$^{-1}$:  so, together the lines indicate a 
velocity dispersion $\sigma \sim 1300$\,km\,s$^{-1}$.    
The weighted mean outflow velocity from the absorption line measurements is 
$\sim 1575\pm254 $\,km\,s$^{-1}$.

We examined the RGS data in the 1-2 keV band to seek confirmation of the 
features detected. RGS data confirm the presence of Mg {\sc xi} and {\sc xii} absorption  lines 
(absence of each is rejected at $p<1\%$ significance), 
yielding 
 line energies $E=1.358^{+0.004}_{-0.002}$,  $E=1.482\pm0.005$ keV with corresponding  bulk velocities 
1239  and  2000 km\,s$^{-1}$  for the outflows. The data are consistent with the detection of 
the Si {\sc xiii} absorption line 
but RGS has so little area at this energy that no useful line constraints could be obtained. RGS has a slightly lower 
energy resolution than the HETG and could not detect the weak emission lines in this regime. 

To put the new results into context, we now fit the broadband pn spectrum 
 utilizing  models for the photoionized gas run from {\sc xstar} 
v21ln \citep{kallman04}. 
As this source is known to show a large degree of flux-related spectral variability we separated the 
pn data into high, mid and low-state spectra, as indicated in Fig.\, \ref{lc}.   
For the initial fit we took  the spectral data  encompassed by the mid-flux state  
and fit across 1-10 keV. Data in this band require inclusion of four layers of gas  in addition to 
a small column of neutral gas fixed at the Galactic line-of-sight value ($N_H=4.08 \times 10^{20} {\rm cm^{-2}}$, \citealt{dl}). Table~2 details  
the results of fitting the pn spectra covering the 
mid-level flux (note that the constraints on the ionization 
parameter of zone 4 were unique in being derived from the 
HEG and not the pn fit).  

The two low-$\xi$ layers of gas show  signatures in the  RGS data below 1 keV and  provide some 
of the broad spectral curvature. 
These two layers (Table 2: zones 1 \& 2)  show a similar total column density to that typically 
detected in the UV band for NGC 3516. No significant features are expected from zones 1 or 
2 above $\sim 1$ keV and as this gas is likely to exist relatively distant from the active nucleus 
we do not discuss those zones in detail in this paper. 

In the pn fit we found all layers of gas to be consistent with full covering of the source except for 
zone 3 (Table~2) whose covering fraction is $45\pm4$\% 
for gas with $N_H \sim 2 \times 10^{23} {\rm cm^{-2}}$, log $\xi \sim 2$. 
The most 
extreme conditions detected are in a layer producing the He-like and H-like Fe absorption lines where the line depths can be 
fit with a zone of column density  $N_{H}=2.62 \times 10^{23} {\rm cm^{-2}}$. 
Overall this 4-zone model gave $\chi^2=2093/1746\ d.o.f$ 
with most of the contribution to $\chi^2$ remaining in the soft-band 
near 2 keV. This may be indicative of a need to refine the model  but as additional features 
indicated by the pn close to 2 keV do not show up in the 
contemporaneous HETG data, and  as 
this is also a region 
most prone to pn calibration  uncertainties we do not attempt to further define  the model based on that part of the spectrum. 

Having found a reasonable solution for the source it is interesting to revisit the individual lines detected in HETG. 
Examination of the {\sc xstar} solution for zones 3 and 4 is particularly interesting. Zone 4 is of such high ionization-state that 
little absorption arises from that gas below the Fe K regime. It is most likely that the lines detected in the 1-3 keV band arise in 
zone 3, the partially-covering absorber. We conclude from the combined HETG and pn results that  
the partially-covering zone 3 gas has an intermediate ionization-state and large column composed of 
 clouds with velocity widths $\sigma \sim 1210\pm 1210 $ km\,s$^{-1}$  and bulk outflow  
velocity  $\sim 1575\pm 254$\,km\,s$^{-1}$. 

A model utilizing an ionized reflector (parameterized using the {\sc reflion} model of Ross \& Fabian 2000) 
in place of zone 3, but allowing in the fit the other three layers of warm absorbers 
 did not fit the mid-state spectrum 
at all well ($\chi^2=3441/1714\ d.o.f.$). While some reflection from the absorbing clouds 
must (at least) be present in this source it appears the absorption effects are dominating the spectral shape and any additional reflection was not discernible in these data. 
{\it XMM} lacks the broad bandpass needed to separate out any reflection component from  the absorbers 
  and reflection-dominated models are not considered further in this paper. 

\subsection{Spectral Variability}

Having confirmed the previously-established picture for NGC 3516 \citep{turner05} as a source dominated by layers of complex 
absorbing gas, we then investigated the spectral variability of the source in the context of such a model. 

First, we attempted to model the spectral steepening with flux by allowing each of the key absorber parameters  
(ionization-state, column density and covering fraction)  for each layer to vary in turn. 
We found only one way in which a variation of a single fit parameter could adequately model 
spectra exhibited over the full flux 
range traversed by the source and that was a change in covering fraction of zone 3. As noted previously, 
the mid-state data represent a covering fraction of $45\pm 4$\% for zone 3. A good fit can be achieved for the 
 high-state spectrum by allowing a simple drop in covering fraction to 
 $37\pm 5$\% and for the  low state by requiring an increase to $67\pm 7$\% for zone 3. 
All other single-parameter explanations for the observed spectral variability could be ruled out 
at $p<0.1\%$ significance (mostly failing to fit the low state). Specifically, 
the popular model where the relative strengths of a variable power-law and steady reflection component change 
cannot account for the deep dip spectrum (this is ruled out at $p<0.1\%$ significance). 
In the context of the variable-covering  model for spectral variability then the 
{\it Suzaku} observation \citep{markowitz07} caught  NGC 3516  hidden by a large cool cloud -  
the overall behavior of NGC 3516 appears to be a mixture of cloud passages with changes in 
ionization-state of the 
absorbing zones. What is perhaps most surprising is the rapidity of some of the changes observed. 

Using crude time-resolved spectroscopy we found that for both the  flare and dip events, 
the change in covering fraction accounts for all but 15\% 
of the source flux variations that are observed, leaving a 
question as to whether all observed variability can be caused by simple absorption changes.  
We therefore constructed a softness ratio  to investigate the spectral variability at 
finer timescales than allowed by  time-resolved spectroscopy. 
Fig.\,\ref{softness} shows the  softness ratio 
(0.5-4.0/4.0-10.0 keV count rate)  versus the total 
0.5-10.0 keV count rate compared to a model line representing 
the values allowed by our multi-zone model, for different values of covering fraction (assuming all other parameter values to be fixed, including intrinsic continuum normalization). 
The model line terminates at the point where the source is completely uncovered. 
 The source behaviour is broadly consistent with the model line 
 in the range range 10-15 ct/s. Below 10 ct/s 
there is some residual soft flux representing some unmodeled soft emission. Above 15 ct/s there are some 
data showing the source to be softer than expected, these data may represent times 
when the ionization-state of 
one of the gas layers has increased (the ionization-state of absorbers laying outside of zone 3 would be 
expected to increase as the covering fraction of zone 3 went down because of changes in the incident 
illuminating flux). Flux changes above 20 ct/s are consistent with no spectral variation and may 
represent intrinsic continuum variability. 
 
\begin{figure}[h]
\resizebox{6cm}{!}
{
  \rotatebox{-90}{
  \includegraphics{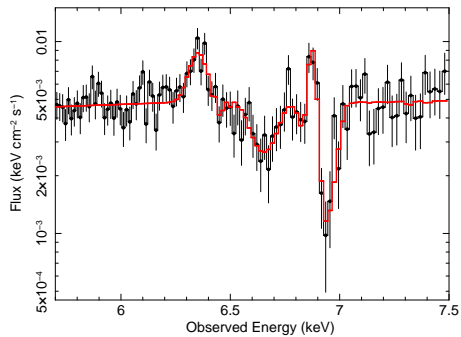}}}
  \caption{Spectral data from the summed HEG first-order data 
from the five data segments taken during Oct 2006. 
The solid line shows the model fit to the data. 
 }
\label{heg}
\end{figure}

\begin{table}
{\small
\begin{tabular}{cr@{}lr@{}l}
Zone & \multicolumn{2}{l}{column density} & \multicolumn{2}{l}{log ($\xi$/erg\,cm\,s$^{-1}$)}  \\ 
 & \multicolumn{2}{l}($10^{22}$\,atoms\,cm$^{-2}$) &  \\
1 & 0&.24$^{+0.03}_{-0.02}$ & $-2$&$.43^{+0.58}_{-0.03}$ \\
2 & 0&.05$^{+0.01}_{-0.01}$ & 0&.25 (f) \\
3 & 20&.2$^{+8.7}_{-3.2}$ & 2&.19$^{+0.07}_{-0.07}$ \\
4 & 26&.2$^{+6.3}_{-8.7}$ & 4&.31$^{+1.19}_{-0.14}$ \\
\end{tabular}
}
\caption{X-ray Absorber Parameters and 90\% confidence errors, fixed parameters are denoted (f)}
\label{scatttable}
\end{table}

\begin{figure}[h]
\resizebox{6cm}{!}
{
  \rotatebox{0}{
  \includegraphics{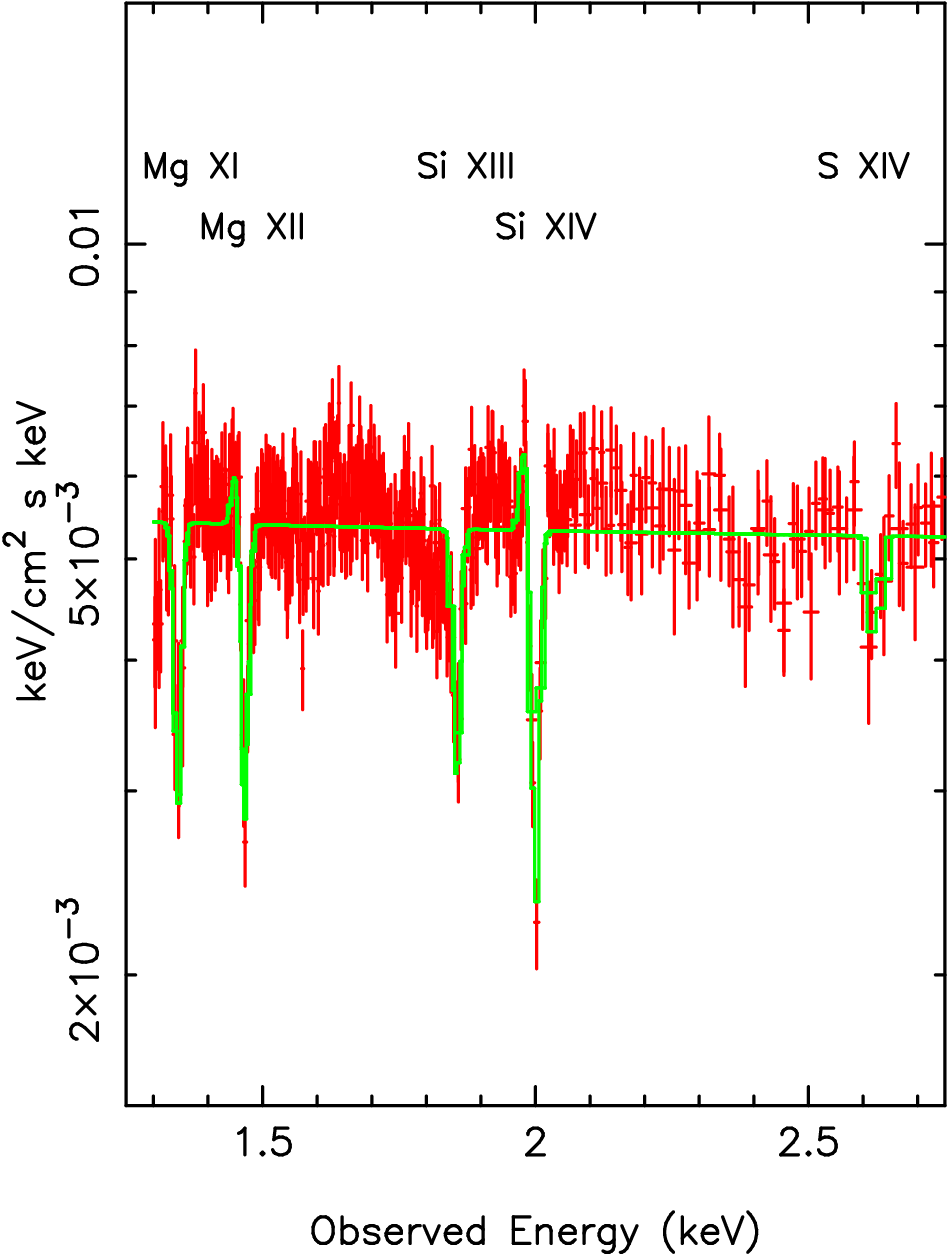}}}
  \caption{Summed  first-order MEG data and summed first-order HEG data 
from the five segments accumulated during Oct 2006. The 
two grating datasets are shown coadded, the solid  line shows the model 
detailed in Table 1.} 
\label{meg}
\end{figure}

Finally, we note the presence of weak transient Fe emission features in the spectra, 
of the type first discovered in 
the 2001 observation of this AGN \citep{turner02}. 
The addition of a narrow ($\sigma \sim 3$eV) gaussian emission line  at 
$E=6.20\pm 0.02$ keV was required in the pn data
(absence of a line being rejected at $p<1\%$ significance),
with line flux  $n=9.4\pm 1.4 \times 10^{-6}$ 
photons cm$^{-2} {\rm s}^{-1}$,  EW=23$\pm 4$ eV. Addition of a second  line to the model  yielded 
$E=5.98\pm 0.03$ keV,  $n=6.0\pm 2.0 \times 10^{-6}$ 
photons cm$^{-2} {\rm s}^{-1}$,  EW=15$\pm 5$ eV.  Some residuals are also evident in the HEG spectrum at those 
energies supporting the reality of the features  (Fig.\,\ref{heg}). 
Interestingly, the features show up most prominently in the 
high-state spectrum (Fig.\,\ref{2006_states}), as they did during the 2002 observation. This 
tendency for Doppler-shifted 
Fe emission lines to show up in source high flux states 
supports a close link to the continuum source flux with little lag, 
ie an inner disk origin. 

\begin{figure}[h]
  \resizebox{8cm}{!}{ 
  \rotatebox{0}{
  \includegraphics{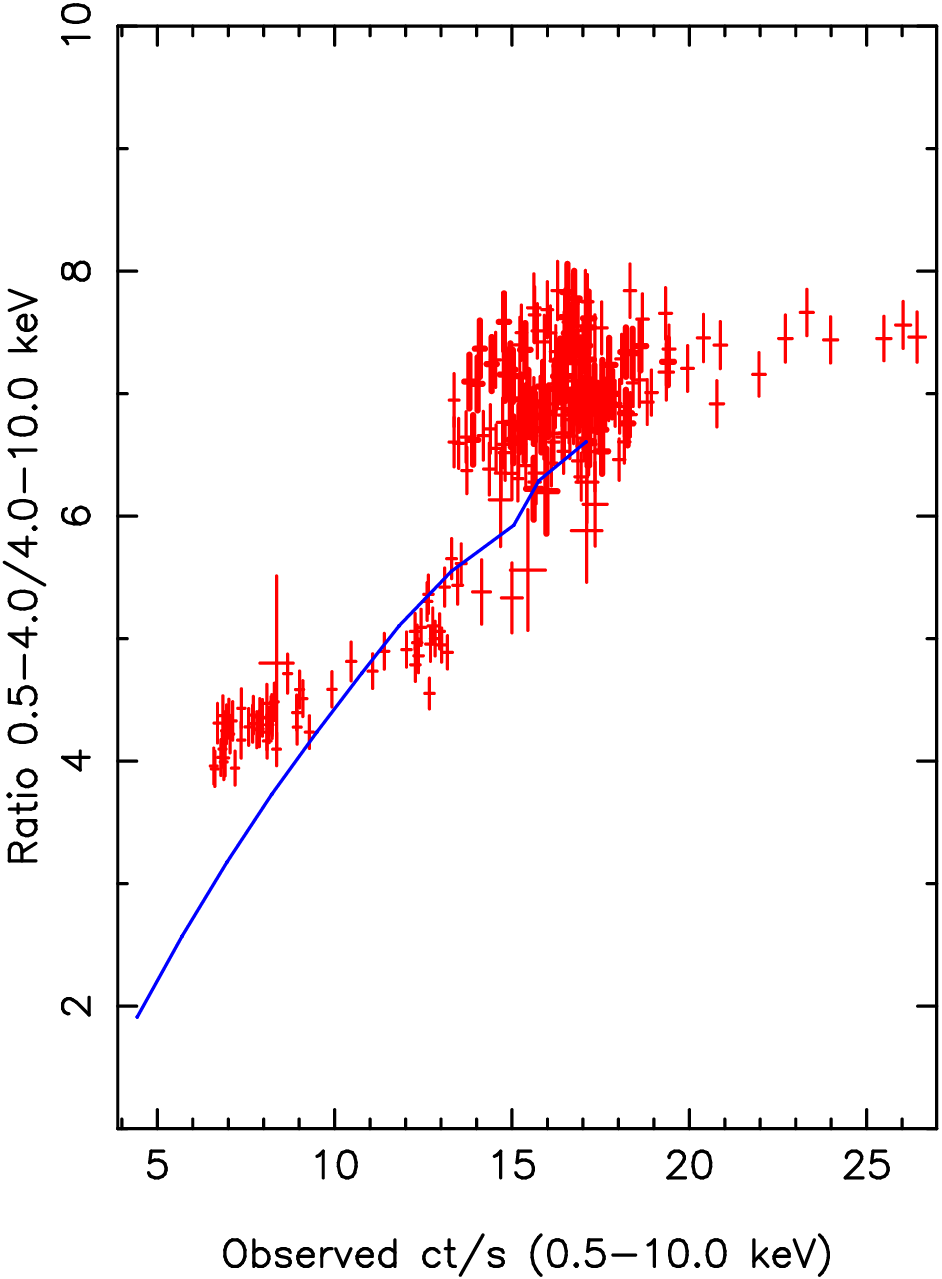}}}
  \caption{The ratio 0.5-4.0 keV/4.0-10.0 keV pn cts on the 
y-axis versus the total pn count rate 0.5-10.0 keV 
on the x-axis. The solid model line shows the changes expected if the 
only variable is the covering 
fraction of zone 3 gas, the model line terminates at the point where none of the continuum source 
is covered by zone 3. 
} 
\label{softness}
\end{figure}

\section{Discussion}

Here we detect four distinct layers of ionized gas absorbing the nuclear continuum. The three zones 
detected during 2001 
have obvious counterparts in 2006 data, and a new, highly ionized layer is detected for the first time 
in this source. 
In our new analysis we also find two low-ionization zones of gas, similar to those known to cause absorption 
in the UV regime, and as these likely 
arise quite far from the nucleus, are not the topic of detailed discussion here. We concentrate on 
the implications of the findings for the innermost zones, 3 and 4. 

\subsection{The Partial-Covering Gas: Zone 3}

The 'heavy' partial-covering zone detected by \citet{turner05} is very similar in all respects to 
the layer denoted ``zone 3'' here: this is a zone characterized by 
intermediate ionization-state (log $\xi \sim 2$), large column density 
($N_H \sim 2 \times 10^{23} {\rm cm^{-2}}$) and modest 
covering fraction, varying around  $\sim 45$\%, similar  to that observed in 2001.
 
The source light curve shows a  deep dip in flux having an   
egress of  duration 
$\sim  30$ ks.  Time resolved spectroscopy indicates that the 
 simplest explanation of the spectral variability observed during the dip 
 is a simple change in covering fraction of the zone 3 absorber. 
 In the context of this model the {\it Suzaku} spectra may be 
interpreted as a  long duration eclipse by a cool gas zone, leading to 
an association of the {\it Suzaku} `primary absorber' 
\citep{markowitz07} with zone 3. 
Examination of the softness-flux plot (Fig.\,\ref{softness}) 
indicates that while covering changes in zone 3 may 
cause deep observed dips in X-ray flux, 
other important phenomena are probably at work to produce the total light curve 
and spectral variations observed in the source. Possibilities 
include  ionization changes 
of the absorber and some intrinsic variation of the source continuum flux. 

The evidence for 
rapid changes in covering fraction of an X-ray absorber places this BLSy1 galaxy in the 
company of sources such as 
 the narrow-line Seyfert 1 (NLSy1) galaxies MCG-6-30-15 \citep{mck},  Mrk 766 \citep{turner07}, 
and the BLSy1 1H0419-577 \citep{pounds04}; all 
of which  have been discussed by some authors in terms of variable absorption. 
Partial-covering models have been raised at various epochs as a possible explanation for 
Seyfert spectral and timing properties  (e.g. \citealt{boll97}), the complete ambiguity between 
absorption and reflection dominated models in low-resolution X-ray spectra available to date 
  left it unclear 
which process was most dominant. However, recent results have shown that 
the role of absorption has not been adequately accounted for and 
consideration of rapid absorption variability is obviously 
key to any  meaningful analysis of the timing and spectral variability properties  
of active galactic nuclei. 

The presence of large amounts of 
optically-thick gas close to the active nucleus means that both absorption and reflection from that 
material must contribute to the observed 
spectrum and in isolating key events, careful decomposition of the data train may lead to 
significant new insight into the nature of the reprocessed emission. 

As noted previously, the detailed shape of the dip event in NGC 3516 is complex (Fig.\,\ref{dip_zoom}). 
Some fraction of the source is completely covered for 10 ks and then the egress begins with a 20 ks 
shallow recovery followed by a rapid rise over 6 ks. The shape of the dip is strikingly similar 
to that observed for a deep dip in MCG-6-30-15 \citep{mck}, indicating that the same physical 
process is likely responsible 
for deep flux dips across the Seyfert population. 
 
If one considers a spherical cloud of gas of constant density occulting a point continuum source, 
although the cloud shape produces some gradation in opacity as the blob traverses the sight-line, 
this crossing would produce a smoothly varying light curve, and symmetric in ingress/egress times, 
rather than the asymmetric curve that is observed. 

As suggested for MCG-6-30-15  by \cite{mck} the dip profile shape 
can be explained by a continuum 
source having non-uniform emissivity. If the emitter is composed of a central bright source embedded in 
a  fainter 
extended  emission region, one could get a dip profile similar to that 
observed as a cloud crosses the line-of-sight. 
 \citet{mck} sketch a schematic of the general picture (their Fig. 2), applicable also 
to NGC 3516 (although we require no gap between the central and extended components of emission).  
The dip profile observed in NGC 3516 is 
consistent with the \citet{mck} scenario.  
Another possibility is that inhomogeneities exist  in the absorber, and that this complex absorber 
crosses a 
simple point-like continuum source. 
The  asymmetry of the ingress/egress timescales gives us some additional information:  
these can arguably be more naturally produced with an inhomogeneous absorber crossing the sight line than 
with a simple cloud crossing a spatially-complex emitter. In the latter case we would need to invoke additional complexity such as that  the central source is asymmetric and that the occulting 
material does not simply 
cross the source but maybe changes trajectory 
 during its passage across the line-of-sight 
egressing across a different spatial dimension of the continuum emitter than 
it ingressed across.  
 Complex situations such as this might arise if a cloud of gas is lifted up off the disk as it orbits,
 ie if the absorber is part of a disk wind.  

HETG data trace individual lines arising in zone 3, 
showing a mean outflow velocity $\sim 1600$\,km\,s$^{-1}$, similar to the outflow velocity  
$\sim 1100$ km\,s$^{-1}$  found for 
what appears to be the same  zone observed during 2001.  If this zone appears to be 
occulting the central emission/reflection regions then the gas must have radial and 
transverse components of motion. 

 As the transverse component of velocity is unknown, we express 
our derived quantities in units 
likely applicable to such sources. Using a mass estimate for the central black hole,  
M$=2.95 \times 10^7 {\rm M}_{\rm \odot}$ \citep{nik06} we obtain 
an estimate of the radial location of zone 3, 
$R \simeq 3.8 \times 10^{15} \, v_4^{-2} {\rm cm}$ where 
$v_4$ is the orbital velocity in units of 
$10^4$ km s$^{-1}$. For a velocity of 10$^4$ km s$^{-1}$  $R \sim 1.5$ light days, 
locating the material on the inner edge of the BLR \citep{wanders93}.
 
The fitted ionization parameter, $\xi$ may be related to 
the radial location of the gas zone, $R$, the gas density $n$ and the (intrinsic) ionizing luminosity 
using  $\xi = L_{\rm ion}/R^2\,  n$ and thus can be used to estimate $n$ (these, and other simple
relationships used here are derived in detail by \citealt{ckg} and \citealt{blustin}).   
The ionizing luminosity utilized by {\sc xstar} is defined 
over the  1-1000\,Ryd band. The observed luminosity is 
$L_{\rm obs} \simeq 1.2 \times 10^{43}$\,erg\,s$^{-1}$;  correcting for the 
fitted absorption layers gives an  implied intrinsic luminosity 
$L_{\rm ion} \simeq 4.8 \times 10^{43}$\,erg\,s$^{-1}$. Utilization of the implied intrinsic luminosity 
gives  $n \simeq 2.1 \times 10^{10}\, v_{4}^4 {\rm cm^{-3}}$. 
Combining this with the 
relation $N_H \simeq n \Delta R $ where $\Delta R$ is the shell thickness, 
we obtain $\Delta R \simeq 9.2 \times 10^{12}  v_{4}^{-4} {\rm cm}$ and 
$\Delta R/R \simeq 2.6 \times 10^{-3}   v_{4}^{-2} $, indicating 
zone 3 to be a thin shell of gas. 

For zone 3, 
 the estimate of radial location and comparison of assumed transverse 
and measured outflow velocity indicates that the 
 gas could  be gravitationally bound. While a transverse 
velocity of order  $\sim 10^4$ km s$^{-1}$ is necessary to obtain a 
reasonable geometry, the value is subject to large 
uncertainty.  Further, 
as we only measure the outflow component along our line-of-sight  
the true outflow velocity may be larger than observed and 
zone 3 is thus consistent with having 
 comparable transverse and outflow velocities. 
In addition to this geometrical consideration we note that steady-state 
wind model solutions exist where there is continuous injection or 
conversion of momentum into the wind, and it is not strictly necessary to 
have a measured outflow velocity that is greater than the local escape 
velocity for an outflow ultimately to be unbound.  Hence we cannot at this 
stage make a definite statement about whether the observed outflow forms 
part of a large-scale wind or whether it indicates a more complex 
gravitationally-bound flow. 

The mass in the absorber can be estimated from 
 M $\simeq 4 \pi R^2 N_H  m_p C_g$ where 
 $m_p$ is the mass of a proton and $C_g$ is the 
global covering fraction. 
The mass in zone 3 is estimated as    
M$ \simeq 0.03\, v_{4}^{-4} C_g\, {\rm M_\odot}$. 
The mass flow rate is then 
 \.{M} $\simeq$ M$/t_c$ \citep{ckg} where $t_c\simeq\Delta R/v $ is the radial travel time and here 
$v$ is the outflow velocity rather than the orbital velocity. As noted above, 
for zone 3 we have 
measured an outflow velocity of 1600 km s$^{-1}$ that is 
 is a lower limit on the 
actual outflow velocity, as it is only the component of velocity that is in 
our line-of-sight. We estimate the  mass flow rate as 
 \.{M}$\simeq 16 v_{1600}\, C_g\, {\rm M_\odot}$\,year$^{-1}$, 
this is quantity that does not depend on the unknown orbital velocity.

Assuming the dip event was caused by a simple cloud, its diameter
$d_{cloud} \simeq v_{orb} t$, where t is taken as the time for the egress,
the strongest constraint within the dip event, so $d_{cloud} \simeq 3
\times 10^{13} v_{4}$ cm. For a velocity of $10^4$ km s$^{-1}$ the lateral
dimension of the cloud ($d_{cloud}$) is then approximately equal to the
cloud depth $\Delta R$ and a reasonable cloud geometry is thus consistent
with the likely orbital velocity. The maximum coverage by the absorber is
$\sim 70\%$ suggesting the continuum to be $\sim 1.2$ times the size of
the cloud on the sky. The light-crossing time for the continuum source is
thus $\sim 10^{3} v_4$ s, consistent with observed variability.

\subsection{The Highly Ionized Gas: Zone 4} 

The highly-ionized zone 4 gas detected here has 
log $\xi \sim 4.3$ and column density 
 $N_H \sim 2.6 \times 10^{23} {\rm cm^{-2}}$, adding to the 
evidence for large columns of highly-ionized gas being important in Seyfert nuclei. 
This gas zone is the origin of the H-like and He-like absorption lines from the K shell of Fe but has little 
signature below $\sim 6$ keV. The 2001 {\it XMM} data are inconsistent with the presence of 
He-like and H-like absorption lines of the same flux.   

The very high ionization and column density, combined with apparent variability on a timescale of years and the 
energies at which these lines are detected disfavors an origin of this gas in the warm-hot intergalactic medium. 
That we have detected similarly extreme zones of gas in the longest observations of 
AGN (Mrk 766, \citealt{turner07}; NGC 3783, \citealt{reeves04}; MCG-6-30-15, \citealt{young}) 
suggests their existence could  be common in active nuclei, these features are below  the limit of 
detectability for  typical short AGN exposures conducted to date. 

An intriguing aspect of this extreme zone  is the apparent observation of both the emission and 
absorption components from  
 H-like Fe in  zone 4. The pair of features appears reminiscent of classic `P Cygni' profiles that 
arise in outflowing shells of gas such as stellar winds. In this case it is 
not clear that the components arise from such a simple spherical shell of gas. Most likely, the gas has some 
more complex geometry (such as a bi-cone). 
Previous claims of  `associated' emission/absorption features  
have been made for the lower-ionization gas zones traced by 
soft-band X-ray features (e.g. \citealt{kaspi02}).  
From the velocity separation of the emission and absorption 
components of  
H-like Fe  the implied outflow velocity of zone 4 
is $\sim 1000$ km\,s$^{-1}$  while line widths 
imply a velocity dispersion 
$\sigma \sim 2124^{+850}_{-255}$ km\,s$^{-1}$  (based on 
the isolated H-like absorption line).

Using the equation for ionization parameter as before we 
estimate a density for zone 4, 
$n \simeq 1.6 \times 10^8\, v_4\,^4 {\rm cm^{-3}}$ and 
$\Delta R \simeq 1.6 \times 10^{15} v_4\,^{-4} $ cm, 
$\Delta R/R \simeq 0.4\,  v_4^{-2}$. The mass in zone 4 is M $\sim 0.04  v_4\,^{-4} C_g M_{\odot}$. 
The measured outflow velocity for zone 4 is $\sim 1000$ km/s and the  
mass loss rate 
\.{M}$\simeq \, 0.06  v_{1000}\, C_g\, {\rm M_\odot}$\,year$^{-1}$.  

Turning to the emission signature of zone 4, the 
normalization of the H-like Fe emission line can be compared to the emission predicted by {\sc xstar} 
for a full shell of gas 
covering the entire ionizing source. This comparison shows the observed line flux to be an order of 
magnitude larger than predicted. 
Taken at face value this leads us to an estimate of the global covering factor (covering factor seen over 
$4\pi$ steradians as viewed by the continuum source) 
to be $C_g \sim 11^{+17}_{-6}$ for zone 4, which is unreasonably large, especially given the 
mass and loss rates estimated above.   

Despite the sizable uncertainty on this covering factor it is interesting 
to consider the ways in which the predicted emission line strength may be brought into line with that observed. 
Since the Fe {\sc xxvi} emission is a resonance line, its strength could be increased by photo-excitation, as
suggested for the resonance lines of H and He-like O, N, and Ne in several other 
Seyferts (e.g. \citealt{sako}; \citealt{armen}). 
However, there will be little enhancement by photo-excitation once the line becomes optically thick 
to the continuum 
radiation. Although there can be greater contribution from photo-excitation if the gas is turbulent, the ionic 
column densities are too large in this case ($\sim 10^{19} {\rm cm^{-2}}$).
For example, if the absorbing gas has a 
micro-turbulence $\sigma \sim 1500$\,km\,s$^{-1}$, then 
the predicted line flux would be twice as large as that calculated in the {\sc xstar} table used here. However,  
this is not enough of an increase to bring the predicted and observed fluxes into consistency. 
Even allowing for the absorption-correction to the intrinsic luminosity incident upon zone 4  
doesn't produce any significant  
 increase in the predicted line flux.  The recombination line depends critically on the fractions of 
 Fe {\sc xxvi} and fully ionized Fe, and there is a limit to how much recombination 
line can be produced in the 
gas. 

Nevertheless, the strength of the Fe {\sc xxvi} line is very sensitive to temperature, 
particularly since it is primarily
formed by recombination. Our models predict that 
when that is the dominant Fe ion, the gas must be in the Compton-cooled regime (\citealt{krol}). 
In this case, the gas temperature depends mostly on the spectral energy distribution of the incident radiation.
A weaker X-ray flux or a stronger low energy flux could result in a significantly lower temperature, hence
a stronger Fe {\sc xxvi} line. For example, the presence of a reserve of low energy photons, such as may occur 
close to the inner edge of the accretion disk, 
can drive the Compton temperature down. 
In conclusion we 
note  that  in NGC 3516 Compton cooling appears to be 
important. 

\section{Conclusions}

Emission and absorption features are detected across the X-ray spectrum of NGC 3516 confirming the 
presence of at least four distinguishable layers of absorption that shape the spectrum. 
Emission and absorption lines are detected from  H-like Fe  revealing the presence of one absorbing layer 
 with extreme properties, log $\xi \sim 4.3$, $N_H  \sim 2.6 \times 10^{23} {\rm cm^{-2}}$ 
outflowing at $\sim 1000$\,km\,s$^{-1}$. 

A lower-ionization zone of gas with  log $\xi \sim 2.2$, 
$N_H \sim 2   \times 10^{23} {\rm cm^{-2}}$ covers just  $\sim 45\%$ of the source. Variations in 
covering fraction of this intermediate layer 
account for much of the spectral and observed flux variability in the source. 
In particular, a deep dip in the light curve can be explained as an eclipse 
by a cloud 
of this gas and the shape of the light curve during occultation reveals that either the cloud or the  emission 
region must be inhomogeneous. 
The remarkable similarities between the deep minima of NGC 3516 and MCG-6-30-15 suggest  that 
these events might be  common in Seyfert galaxies. It appears that 
rapid variations in absorber parameters play a much greater role in 
shaping the observed X-ray variability in AGN than has been appreciated to date. These occultation 
events may offer  the best way forward for probing the details of the innermost regions of AGN. 

The detection of a significant outflow velocity for the most 
highly-ionized gas zone in NGC 3516   
 suggests that at least this component of  X-ray absorption  
originates in a wind. 
The small implied radial location for  the highly-ionized   
gas rules out the putative  
torus as an origin for the wind,  leaving a disk wind as the most 
likely explanation.  
\citet{ruiz05} suggest, from optical kinematic studies, that 
NGC 3516 appears to be viewed at a fairly high inclination, ie close to the plane of the 
absorbing material and accretion disk system, similar to NGC 4151 \citep{cren2000}. 
Sources viewed  at shallow angles to the plane of the disk would be  
 the best candidates in which to view disk winds, and  this may have 
contributed to the easy detection of 
outflows in highly-inclined sources such as these.

\section{Acknowledgments}
Based on observations obtained with {\it Chandra}, operated 
by SAO on behalf of NASA, and with {{\it XMM-Newton}}, 
an ESA science mission with instruments and contributions 
directly funded by ESA Member States and NASA. 
TJT acknowledges NASA grants NNX06AH97G and G07-8092X.  
We thank Jianning Zeng for help with the {\it Chandra} data 
reduction and the anonymous referee for comments that improved 
the paper.

\label{lastpage}

\end{document}